\documentclass[pdflatex,sn-mathphys]{sn-jnl}

\jyear{2021}%

\theoremstyle{thmstyleone}%
\theoremstyle{thmstyletwo}%

\theoremstyle{thmstylethree}%

\begin{document}

\title[Inertia Tensor and Machine Learning]{Detection of Alzheimer’s Disease using MRI scans based on Inertia Tensor and Machine Learning}


\author[1]{\fnm{Krishna} \sur{Mahapatra}}\email{ritamahapatra8556@gmail.com }

\author*[1]{\fnm{Selvakumar} \sur{R}}\email{rselvakumar@vit.ac.in}

\affil[1]{\orgdiv{Department of Mathematics}, \orgname{Vellore Institute of Technology}, \orgaddress{\street{Vellore}, \state{Tamil Nadu}, \country{India},
\postcode{632014}}}


\abstract{Alzheimer's Disease is a devastating neurological disorder that is increasingly affecting the elderly population. Early and accurate detection of Alzheimer's is crucial for providing effective treatment and support for patients and their families. In this study, we present a novel approach for detecting four different stages of Alzheimer's disease from MRI scan images based on inertia tensor analysis and machine learning. From each available MRI scan image for different classes of Dementia, we first compute a very simple $2 \times 2$ matrix, using the techniques of forming a moment of inertia tensor, which is largely used in different physical problems. Using the properties of the obtained inertia tensor and their eigenvalues, along with some other machine learning techniques, we were able to significantly classify the different types of Dementia. This process provides a new and unique approach to identifying and classifying different types of images using machine learning, with a classification accuracy of ($90\%$) achieved. Our proposed method not only has the potential to be more cost-effective than current methods but also provides a new physical insight into the disease by reducing the dimension of the image matrix. The results of our study highlight the potential of this approach for advancing the field of Alzheimer's disease detection and improving patient outcomes.}

\keywords{Alzheimer's Disease, Magnetic Resonance Imaging, Inertia Tensor, Machine Learning.}


\maketitle

\section{Introduction}\label{sec1}
Alzheimer's disease (AD) is a progressive neurodegenerative disorder that is characterized by a decline in cognitive and behavioral abilities \cite{alroobaea2021alzheimer}. It is the most common cause of dementia among older adults and its prevalence is expected to increase as the population ages \cite{world2021global}. 

\begin{wrapfigure}{1}{0.3\textwidth}
\includegraphics[width=1\linewidth, height=1.2\linewidth]{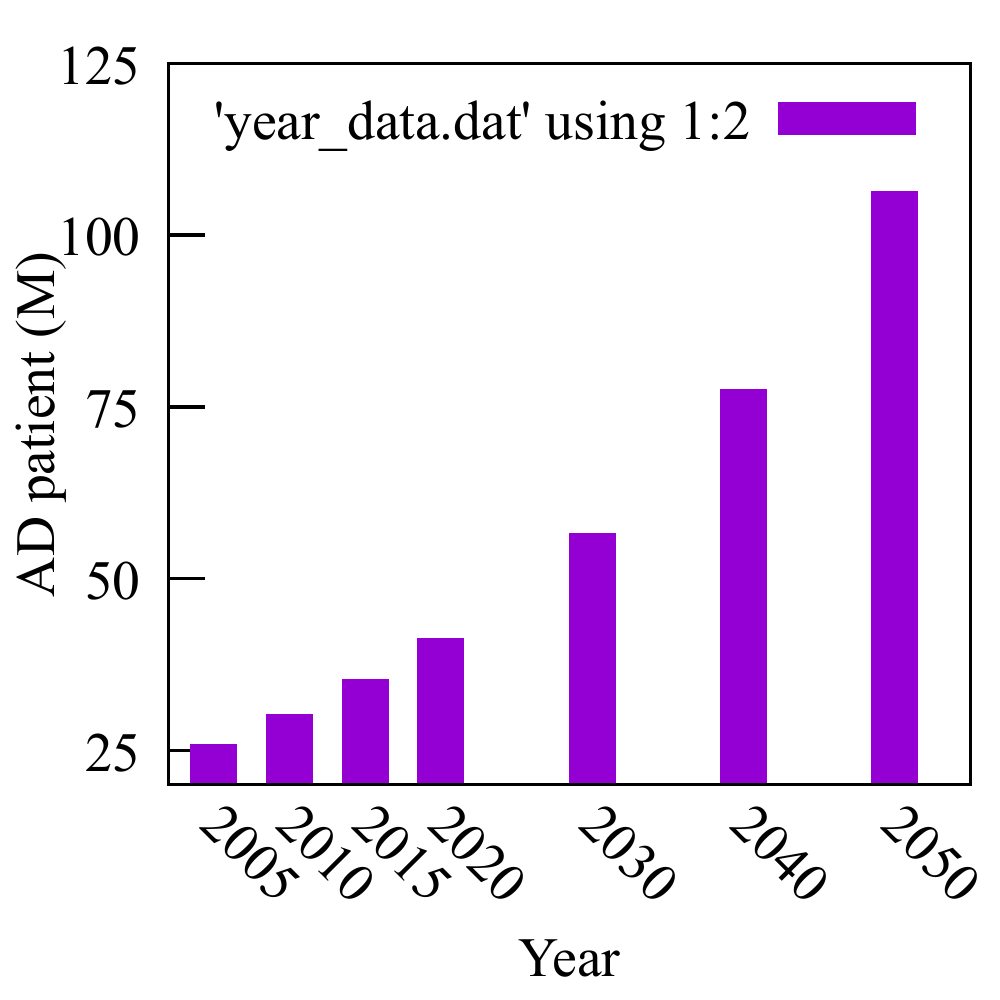} 
\caption{Projected Growth of 
 Alzheimer's Disease Worldwide: 2005-2050.}
\label{fig:wrapBrain}
\end{wrapfigure}

In 2006, approximately twenty-seven million people suffered from this disease worldwide \cite{brookmeyer2007o1}. AD is predicted to affect 1 person in 85 people globally by 2050, and at least $43\%$ of prevalent cases need high-level care for this disease \cite{revi2020alzheimer} Figure\ref{fig:wrapBrain}. Although, the cause of AD is not completely understood \cite{murphy2021mortality} so far.Two kinds of AD are considered, one is sporadic and another one is familial. The sporadic AD ($~ 90-95\%) $ is genetical and environmental and late onset \cite{piaceri2013genetics}. The risk factors of sporadic AD are i) age (1\% for age 60-65, 50\% for age over 85) and ii) gene ($-$allele of apolipoprotein E gene, or APOE-). The familial AD (~5-10\%) is early onset and some inherited dominant gene  speed up the progression of the disease. There is no proven and modifying treatment for AD, a progressive, irreversible brain disorder characterized by a decline in cognitive science \cite{al2020alzheimer}. Therefore, the detection of Alzheimer's disease (AD) at an early stage is crucial for the management and treatment of the disease as well as for planning for the future care of patients and their families \cite{gaugler20222022}. 

Techniques for brain imaging can be used to non-invasively see the pharmacology, function, or structure of the brains \cite{spinelli2003new}. Non-invasive neuroimaging techniques, including Magnetic Resonance Image (MRI), Positron Emission Tomography (PET), and MRI biomarkers, can be used to deduce the clinical mechanisms of AD. However, the cost and complexity of these tools make them difficult to use in low-resource environments \cite{chedid2022development}. There is a significant effort being made to develop strategies when treatment may be most effective,  critical goals remaining early detection of Alzheimer's Disease (AD), particularly at pre-symptomatic stages, to reduce or halt disease progression \cite{nelson2015slowing}.
To accomplish this, cutting-edge neuroimaging methods such as magnetic resonance imaging (MRI) and positron emission tomography (PET) have been developed and utilized to identify structural and molecular changes related to AD \cite{chevignard2020pediatric}.
Additionally, computer aided machine learning methodologies, such as convolutional neural networks (CNNs), support vector machines (SVMs), and logistic regression (LR) (but it performed high prediction for heart diseases) \cite{ambrish2022logistic}, have been increasingly used for integrative analysis in the early detection and classification of AD \cite{lazli2020survey}.
However, the most of studies on Alzheimer's disease that have been published so far have classified it using intelligent classification methods, such as convolutional neural network (CNN) architecture or graph neural network (GNN) architecture. 

In recent years, healthcare has increasingly been using learning-based computer-aided detection systems, such as supervised machine learning (as Naïve Bayes (NB), Decision Trees (DT), K-Nearest Neighbour (KNN)) in the diagnosis of several diseases, including breast cancer \cite{kumar2019machine}, heart diseases
\cite{barik2020heart}, and others. These computer-based studies aim to identify specific characteristics in MRI images that can be used to classify different types of AD \cite{mohsen2018classification}.
Lu et al. (2018) took a data set from 1051 subjects and obtained an accuracy value of 82\% by using a multiscale deep neural network structure to detect Alzheimer's disease \cite{lu2018multiscale}. 
Zhao et al. in their study, used 15 healthy and 15 patient data and obtained 92\% accuracy by using the SVM method\cite{zhang2015detection,zhao2015deep}.
Goo et al. (2017) applied CNN architecture in their study and obtained accuracy values of 87.62\% \cite{gao2017classification}.
Moradi et al.  stated that they used 10 folders for cross-validated and achieved 90.2\% accuracy using SVM and that played a major role in the detection of AD\cite{moradi2015machine}. 
Salvatore et al. in their study for the diagnosis of AD and obtained accuracy rate of 76\% using a machine learning method that optimized\cite{salvatore2015magnetic}.
Das et al. (2013) used least-square SVM, and consisting of 66, 160 and 255 images, and their 5 × 5 CV showed high classification accuracy (on an average $>$ 99\%)\cite{das2013brain}. 
There are several drawbacks associated with using convolutional neural networks (CNNs) for the classification of images, including:  

(1) High computational requirements: CNNs are computationally intensive and require large amounts of data and processing power, making them difficult to implement in some settings. 

(2) Lack of interpretability: CNNs are often considered "black box" models, as it is difficult to understand how the model is making its decisions. This makes it hard to understand what features the model is using to classify the images. 

In this study, we propose a novel method for reducing the dimensionality of brain images to improve the efficiency of the classification process. Specifically, we map each image pixel value matrix to a 2x2 matrix by utilizing the information that, in cases of dementia, brain size may be compressed or expanded. By reducing the image matrix to a 2x2 matrix, the computational cost is significantly reduced. The details of the methods and techniques used are described in section \ref{section: Model Building}.

\section{Data}\label{sec2}
\begin{figure}[ht] \label{MRI_DATA}
    \centering
    \includegraphics[width=\linewidth]{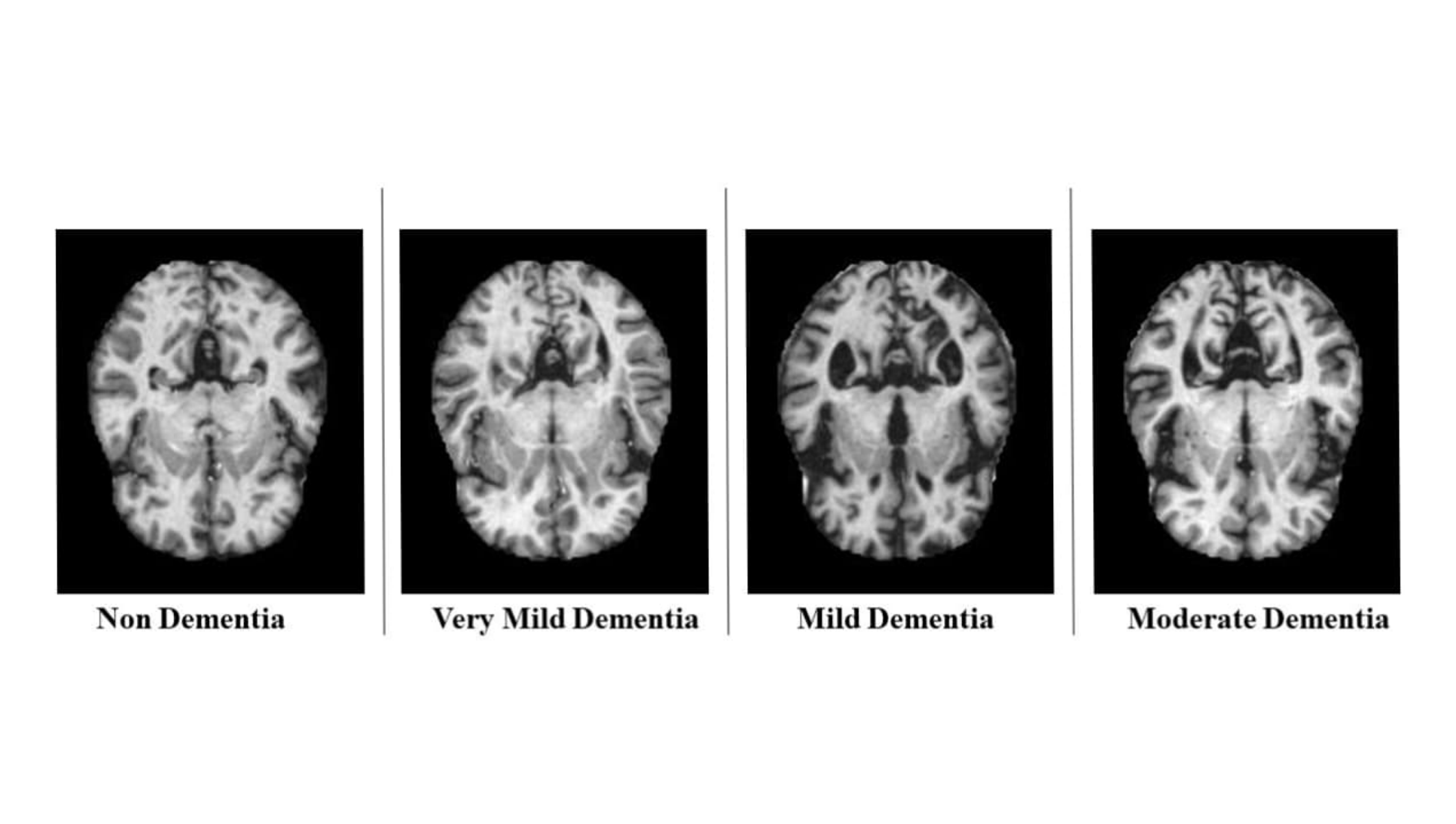}
    \caption{Four Different Class Image}
    \end{figure}

From the open-access Kaggle website \url{https://www.kaggle.com/datasets/tourist55/alzheimers-dataset-4-class-of-images}, the data set was used for this study. The Alzheimer's Dataset consisting of MRI images used has 4 categories:Very Mild Demented, Mild Demented, Moderate Demented, and Non-Demented shown in 
Figure \ref{MRI_DATA}.

\section{Model Building:} \label{section: Model Building}

There are two types of MRI data, one is structural MRI and other one is functional MRI. The most used MRI for AD cases is structural MRI. The structural MRI offers images which provides information about the brain structures, including neurons, synapses, cells, etc. These images helps to deal with the structure of the nervous system and the diagnosis of large-scale intracranial disease (such as the tumor) and injury \cite{Baltes2010}. It measures brain volume in vivo to detect brain degeneration (loss of tissue, cells, neurons, etc.) which is an inevitable progressive component of AD. This work is mostly based on structural MRI images.

A mathematical model of the brain shows how the brain is structurally changing for dementia. In this study, we used two different types of CDR (Clinical Dementia Rate) for two different classes of dementia: subjects with a CDR of 1 were classified as non-dementia, and CDR of 0 as moderate dementia \cite{zhang2015detection}. 

The method of our work is as follows. The basic idea of identifying Dementia from an MRI image comes due to the shrinkage of the brain cell in the images. Note that in the images of MRI scan (which is nothing but a collection of pixel values in a 2-dimensional plane), due to the shrinkage, the distribution of the pixel values in the image will be intrinsically different for different classes of Dementia. Understanding the distribution of the pixel values is therefore key to the detection of Dementia. From the concepts of physics, we see that the Moment of Inertia (MI) tensor is something that helps in understanding the distribution of mass in an extended object. Therefore, by drawing a correspondence of mass of a 2D extended object with the pixel values of the MRI scan image, we can generically calculate the MI tensor of the image itself, which in turn will provide information about the distribution of the pixel values in the image.  
Moving forward, we consider the pixel values in the MRI image $(a_{pq})$ to be equivalent to the mass $(m_i)$ of a 2D object. This assumption simplifies the calculation of the moment of inertia (MI) tensor, which is a $2 \times 2$ matrix for 2D systems. To compute the MI tensor, we first need to define a coordinate system with the origin at the center of the image and the x and y axes aligned horizontally and vertically, respectively. We simplify calculations by assuming that the extent of the 2D object is from $(-1,1)$ in the x and y directions, in some arbitrary unit. It is important to note that these scaling factors and the pixel value to mass conversion factor, which we take to be 1, are chosen for computational convenience and that alternative scaling and conversion factors are unlikely to significantly affect the results.

With the mass distribution, coordinate system, and dimensions of the 2D object defined, the elements of the MI tensor can be calculated as follows:

\begin{equation}
    I = 
    \begin{bmatrix}
    I_{00} & I_{01} \\
    I_{10} & I_{11}
\end{bmatrix}
\end{equation}

where the different elements of the matrix are defined as follows:

$I_{00} = \sum_i m_i.y_i^2$

$I_{01} = \sum_i -m_i.x_i.y_i$

$I_{11}= \sum_i m_i.x_i^2$

$I_{10} = I_{01}$

where the sum is over all the masses in the 2D system, which is basically the pixel values in the image. $m_i$ is the $i-th$ pixel value in the image. $x_i$ and $y_i$ are the $x $ and $y$ components of the position vector $\Vec{r_i}$ of the $i-th$ point in the image.

\begin{figure}[ht] \label{fig:scatter}
    \centering
    \includegraphics[width=10cm]{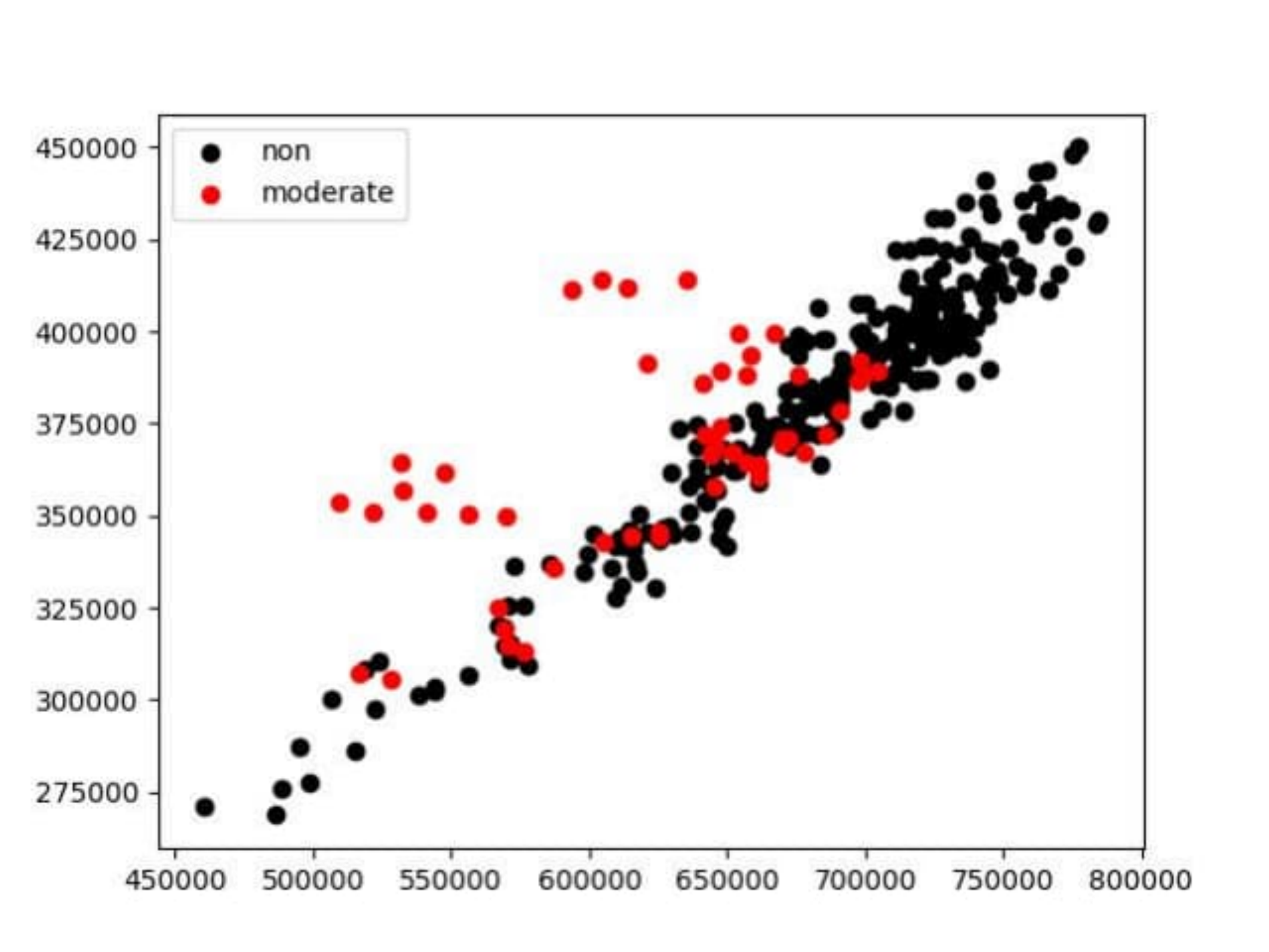}
    \caption{Scatter Plot of Eigen Values}
\end{figure}

Therefore this technique treats each image as a 2D mass distribution and computes the MI tensor for each of them. We then move on to calculate the eigenvalues for each of the MI tensor. We then plot the eigenvalues $\{(I_1,I_2)\}$ ($I_1,I_2$ are the two eigenvalues for each MI tensor) in to observe that the two cases of Non-dementia and Moderate dementia can be distinguished from such plot Figure \ref{fig:scatter}. Using this simple technique of converting an image matrix a $2 \times 2$ tensor and extracting its eigenvalues, we have carried on further analyses to detect different classes of dementia, which are reported in section \ref{sec3}.






\section{Results}\label{sec3}
\subsection{Eigenvalue distribution}

The eigenvalues of the corresponding MI tensor for each of the images related to the known cases of Non-dementia and Moderate dementia is plotted in figure \ref{fig:scatter} which shows a wider distribution of the points for the moderate cases, where a narrower distribution (points mostly following in a straight line) for non-dementia cases. This distribution of the eigenvalues hence can give rise to important classification parameters detecting the different types of dementia.

\subsection{Asymmetry in eigen values}

\begin{figure}[!htb] 
    \centering
    \includegraphics[width=10cm]{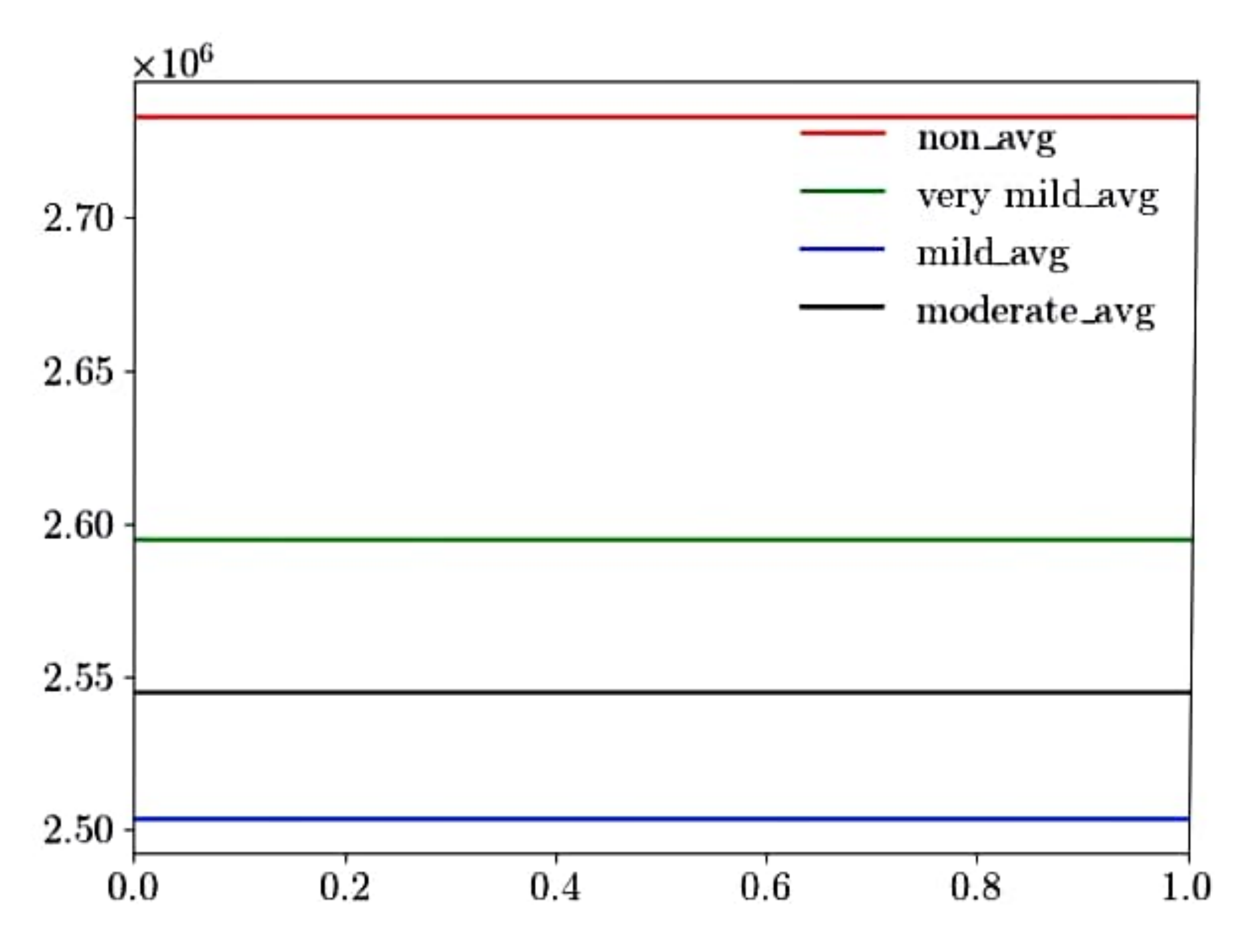}
    \caption{The average value of the asymmetries in the eigen values for each class.}
    \label{fig:asymm_avg}
\end{figure}


The physical significance of the difference between the eigenvalues of the MI tensor is noteworthy as it reflects the amount of mass distributed along each axis. A significant difference between two eigenvalues indicates a greater mass far away from the axis corresponding to the eigenvector of that eigenvalue. As the shrinkage of brain cells in Moderate Dementia alters the pixel value distribution in MRI images, the mass distribution also changes, which is captured by the difference of the eigenvalues (see Figure \ref{fig:scatter}). Our results show that the average difference between the largest and smallest eigenvalues, for all datasets, decreases with the severity of the disease (Nondementia $>$ very mild $>$ mild $>$ moderate). This technique may aid in detecting dementia classes for unknown data, as the higher average asymmetry of eigenvalues may indicate non-dementia.

The reason for this trend can be explained as follows: In the case of non-dementia, the brighter region in the image (section \ref{MRI_DATA}) is larger than in other cases, representing regions of higher mass as their pixel values are larger. Conversely, darker regions denote regions of lower mass as their pixel values are lower. In contrast, for the moderate dementia class, a more "dark" region is present, representing a hollow mass distribution. It can be proven that the asymmetry in the mass distribution (thus the asymmetry in the eigenvalues of the MI tensor) along the principle axes would be larger for solid mass distribution compared to hollow mass distribution, given all other parameters are the same. Therefore, the non-dementia class (equivalent to a solid mass distribution) is expected to exhibit a larger asymmetry or difference in the eigenvalues compared to the moderate dementia class (hollow mass distribution), which is supported by the trend shown in Figure \ref{fig:asymm_avg} as the mass distribution becomes increasingly hollow in the given order.

\subsection{Total pixel value}

\begin{figure}[!htb] 
    \centering
    \includegraphics[width=10cm]{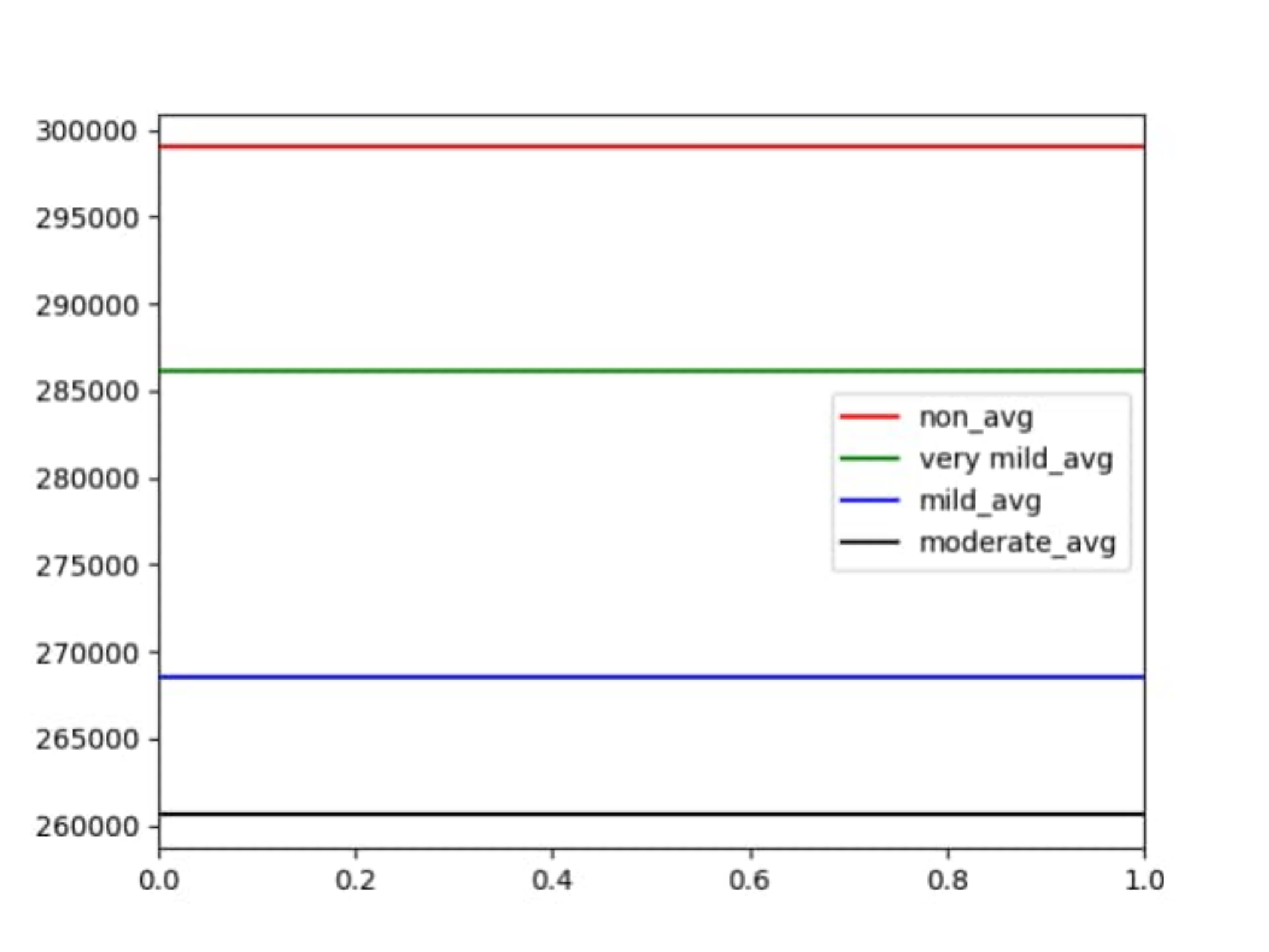}
    \caption{The average values of total mass for each image in various classes.}
    \label{fig:scatter_sum}
\end{figure}

As previously mentioned, we have observed that the MRI images corresponding to the Moderate dementia class exhibit a hollower mass distribution in a physical sense, and that the hollowness increases with the severity of the disease ($H_{non} < H_{very-mild} < H_{mild} < H_{moderate}$, where $H$ denotes the hollowness and the subscripts denote the different disease classes). Consequently, a more fundamental classification scheme can be based on the total mass of the system. We therefore computed the total mass (i.e., the sum of the pixel values) for each image in the various classes and found that the average values of the total mass are consistent with our expectations. Mathematically, since $H \propto 1/M$, it follows that $M^a_{non} > M^a_{very-mild} > M^a_{mild} > M^a_{moderate}$ (see Figure \ref{fig:scatter_sum}), where $M$ and $M^a$ respectively denote the total mass and the average of the total mass for the images, and the subscripts denote the different disease classes. This trend is also illustrated in Figure \ref{fig:scatter_sum}. Based on this simple analysis, we can gain insights into the underlying disease type of unknown image data by comparing the total pixel values with the average value. This analysis will be explore briefly using machine learning methods. There we can use different model to train and test of different classes of asymmetry in eigen values and get the result of classification. 

\section{Machine learning methods}
Prior research has used machine learning techniques to differentiate between Alzheimer's disease patients' images and healthy MRI images. These techniques include decision trees, support vector machines, convolutional neural networks, linear discriminant analysis, multi-regression linear analysis, and others. However, the pre-processed image; features are extracted and given to the classifier as input. The numerous features of class data are explained, and the use of computationally intensive methods that obstruct explainability, such as Inertia Tensor and: black box; ML  models used in this work. Image classification can be accomplished by using feature extraction techniques to reduce a large input data set into important key features. 
\subsection{Feature Extraction:}
\begin{figure}[ht]
    \centering
    \includegraphics[width=8cm]{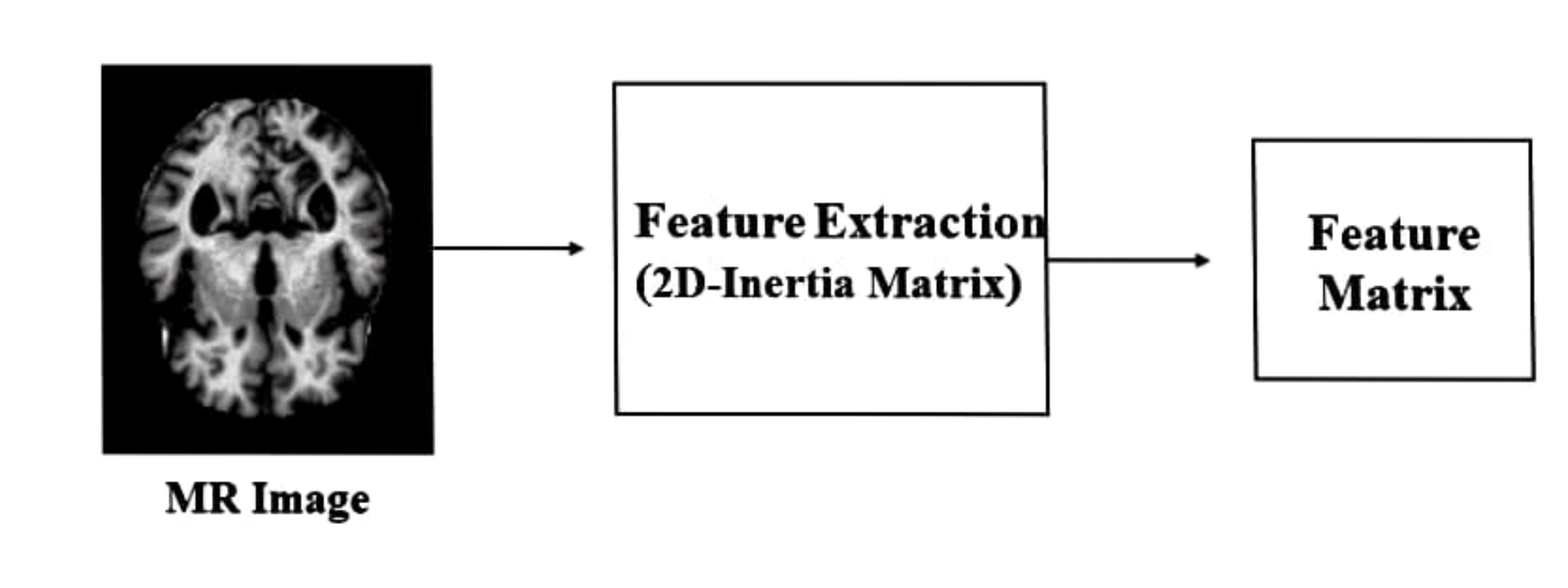}
    \caption{Feature Extraction Process}
    \label{fig:featureMatrix}
\end{figure}

In this study, simplification of the original input data space in order to reduce the original dimensionality into 2D- inertia matrix and obtain a new simpler representation that includes the largest part of the original data variability.Every feature vector of MRI brain images characterizing the images has to be extracted to get exact features. The inertia Tensor Matrix is used as a feature in the design of the mathematical model Figure \ref{fig:featureMatrix}.

\subsection{ML-Based Classification Techniques:}

We now give a brief overview of the machine learning (ML) classifiers that enable us to differentiate between the four different classes of Alzheimer’s disease. We are classifying every dementia class eigenvalue. These ML techniques are known as supervised classifiers. In this paper, we have used one supervised classifiers,i.e., SVM (Support Vector Machine). Support vector machines (SVM), a mathematical model extension of neural networks, can classify both linear and non-linear data. By transforming the input training data into higher dimensional space and constructing hyperplanes (nothing but decision planes). Different kernel functions, such as linear, polynomial, and gaussian
radial basis functions, can be used to specify the decision boundary function. It is a supervised learning
technique that is most employed for regression, classification. One of
the best classification techniques in machine learning, SVM, has been arguably acknowledged \cite{zhang2012classification}. SVMs are linear classifiers, but for non-linear data, we introduced the kernel SVMs (KSVMs), RBF kernel. So, Predictions can be made using machine learning classifiers that the algorithm can predict the class the data
belongs to.

\subsection{Inertia tensor method}
\begin{figure}[ht]
    \centering
\includegraphics[width=\linewidth]{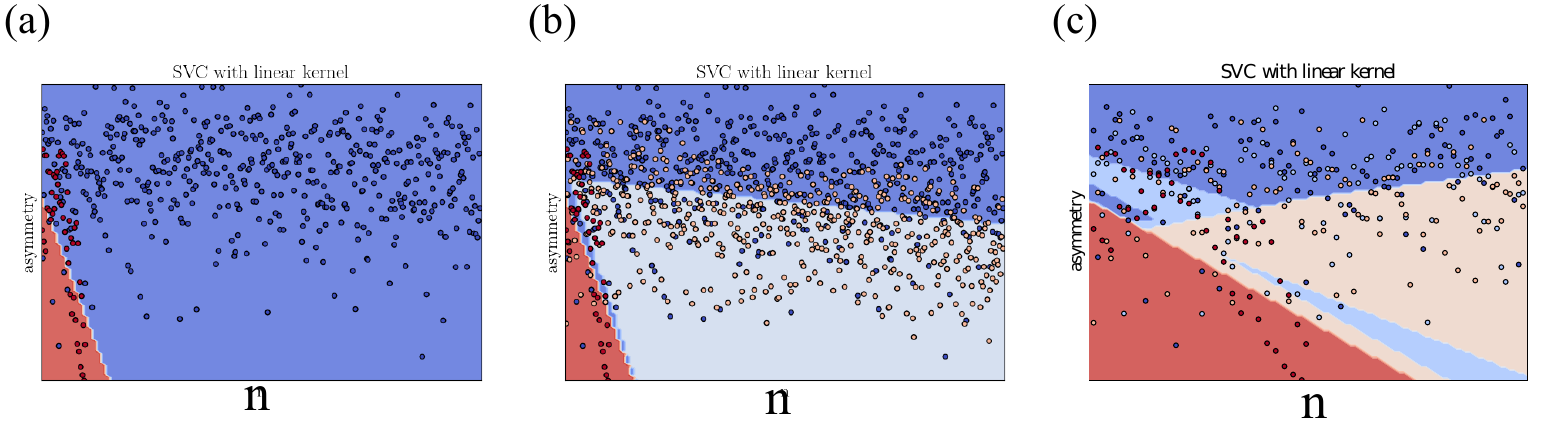}
    \caption{(a)SVM classification of moderate and non dementia images (b) SVM classification of moderate, mild, and non dementia images (c) SVM classification of moderate, mild, very mild, and nondementia images}
    \label{fig:Data}
\end{figure}

As has been shown in the results of section \ref{sec3}, the novel method of creating an inertia tensor from an image by taking the pixel values as masses of a 2D body and scaling the figure dimension to the length and width of the body, which makes it possible to identify an MRI scan image with a physical 2D mass distribution of a body, can also help in differentiating or classifying the different classes of dementia. The technique is summarised in Figure \ref{fig:featureMatrix}. The figure depicts the initial distribution of pixel values in the image. Then by appropriate choice of origin and axes system, we create a $2 \times 2$ inertia tensor. Then we calculate the eigenvalues, which essentially are the same as the diagonal elements of the inertia tensor after diagonalization. Then from the eigenvalues themselves, and their asymmetries, we have shown that classification can be done for the different classes of dementia fig \ref{fig:Data}. The calculation of total mass also is an effective tool for classification. 

\begin{figure}[ht]
    \centering
\includegraphics[width=8cm]{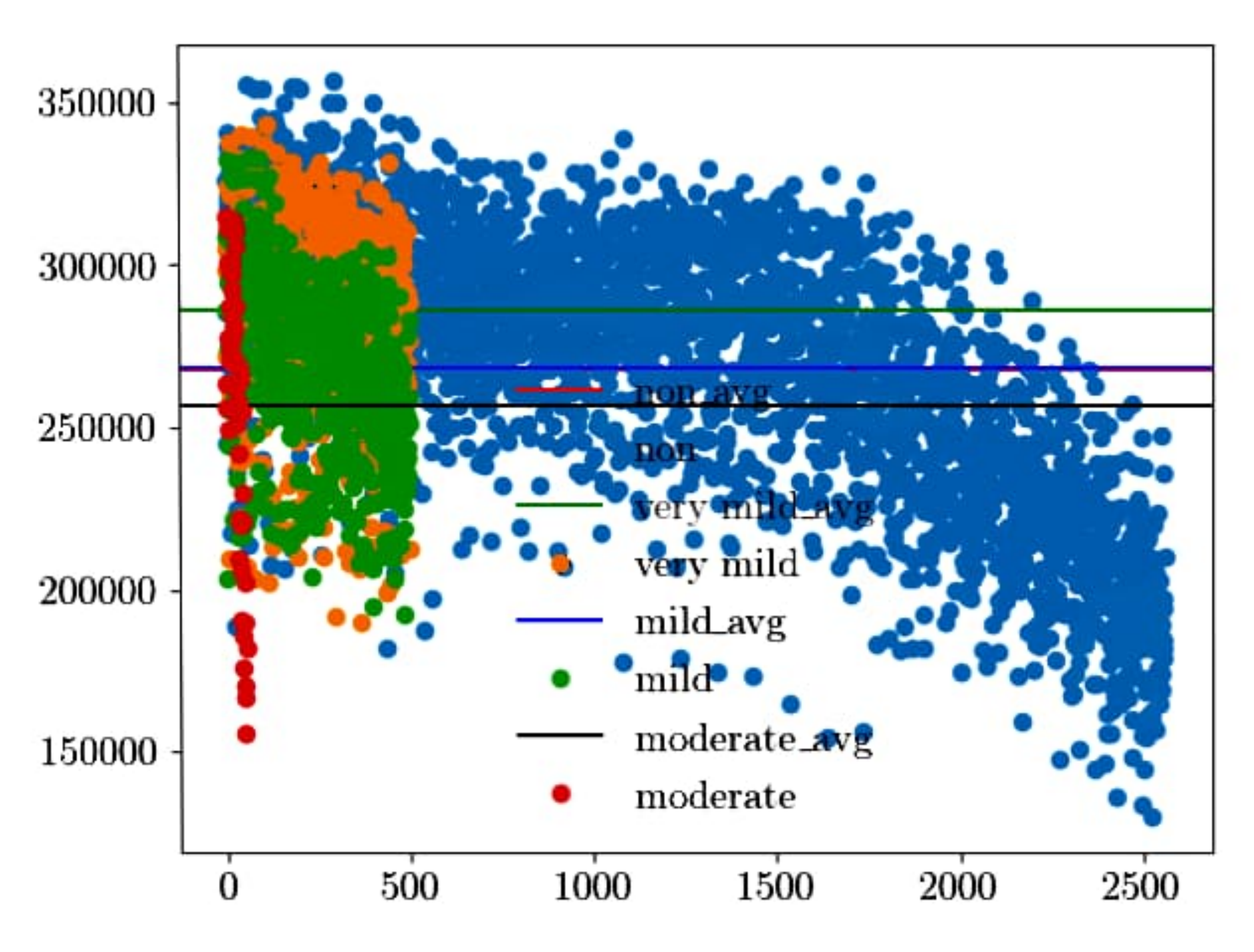}
    \caption{All Asymmetry Class}
    \label{fig:Asymmetry}
\end{figure}

\section{Conclusion}\label{sec13}
This study presents a new method for detecting Alzheimer's disease using MRI scans and machine learning. The approach draws a correspondence between a 2D image and a 2D mass distribution of a physical body and then uses inertia tensor analysis from that mass distribution, calculates the eigenvalue of the inertia tensor matrix, and classify different types of images. With a classification accuracy of ($90\%$), this method has the potential to be more cost-effective and provide physical insight into the disease. The results demonstrate the potential of this approach for advancing Alzheimer's disease detection and improving patient outcomes.

\section{Conflict of Interest}
There is no conflict of interest.

\section{Data Availability}
\url{https://www.kaggle.com/datasets/tourist55/alzheimers-dataset-4-class-of-images}.

\backmatter

\bigskip


\bibliography{sn-bibliography}


\end{document}